\documentclass[12pt,preprint]{aastex}

\shorttitle{Subclasses of SDSS Unclassified Objects}

\begin{document}
\title{Objective Subclass Determination of Sloan Digital Sky Survey
  Spectroscopically Unclassified Objects} 
\author{David Bazell}
\affil{Eureka Scientific, Inc., 6509 Evensong Mews, Columbia, MD
  21044}
\author{David J. Miller}
\affil{Department of Electrical Engineering, The 
Pennsylvania State University}
\author{Mark SubbaRao}
\affil{Department of Astronomy and Astrophysics, The University of
  Chicago}

\begin{abstract}
We analyze a portion of the SDSS photometric catalog, consisting of
approximately 10,000 objects that have been spectroscopically
classified into stars, galaxies, QSOs, late-type stars and unknown
objects (spectroscopically unclassified objects, SUOs), in order to
investigate the existence and nature of subclasses of the unclassified
objects.  We use a modified mixture modeling approach that makes use
of both labeled and unlabeled data and performs class discovery on the
data set.  The modeling was done using four colors derived from the
SDSS photometry: $(u-g)$, $(g-r)$, $(r-i)$, and $(i-z)$.  This technique
discovers putative novel classes by identifying compact clusters that
largely contain objects from the spectroscopically unclassified class
of objects.  These clusters are of possible scientific interest
because they represent structured groups of outliers, relative to the
known object classes.  We identify two such well defined subclasses of
the SUOs.  One subclass contains 58\% SUOs, 40\% stars, and 2\%
galaxies, QSOs, and late-type stars.  The other contains 91\% SUOs,
6\% late-type stars, and 3\% stars, galaxies, and QSOs.  We discuss
possible interpretations of these subclasses while also noting some
caution must be applied to purely color-based object classifications.
As a side benefit of this limited study we also find two distinct
classes, consisting largely of galaxies, that coincide with the
recently discussed bimodal galaxy color distribution.
\end{abstract}

\keywords{galaxies: general --- methods: data analysis --- methods:
 statistical --- surveys}

\section{Introduction}
One of the main goals of modern photometric and spectroscopic surveys
is to understand and isolate the different types of measured objects.
This is best done automatically, both due to the overwhelming number
of objects in the surveys and because of the relative objectivity
gained by using automated methods.  There are a number of approaches
to classification of objects, traditionally falling into two major
groups: supervised and unsupervised.  Supervised classifiers make use
of labeled training data in order to train the software to recognize
certain patterns in the data features that are characteristic of the
different object classes known to be present.  Unsupervised classifiers,
or clustering algorithms, try to find natural groupings of objects in
feature space.  \citet{Storrie92} were the first to use neural
networks, a form of supervised classification, to automatically
classify galaxies according to their morphology.  \citet{Odewahn92}
have developed successful neural network based methods for star/galaxy
separation.  Since the mid-1990s there has been a lot of work using
neural networks for stellar object classification.  \citet{Naim95a, Naim95b} used
backpropagation neural networks to classify galaxies based on their
morphology and showed the results compared favorably to classification
by humans.  \citet{Vonhipple94} and \citet{Bailer-Jones97,
Bailer-Jones98} used neural networks to successfully predict effective
temperatures, MK classifications, and luminosity classes of stars
based on stellar spectra.  \citet{Ball04} used a neural network and a
variety of morphological and photometric parameters to predict the
eClass of SDSS galaxy spectra.  The eClass is a continuous
parameterization of the galaxy type derived from principal component
analysis applied to a collection of SDSS galaxy spectra \citep{Connolly95,
Connolly99}.  \citet{Willemsen05} determined the metallicities of
stars from two different globular clusters using neural networks.

Another supervised classification method that has been used with
astronomical data is the decision tree inducer.  \citet{Weir95a} used
a decision tree to classify stars and galaxies from the Palomar Sky
Survey (DPOSS) based on a set of eight features.  \citet{Owens96} used
an oblique decision tree classifier (OC1) to study the same data as
\citet{Storrie92} used with a neural network, achieving comparable
results.  \citet{Jarrett00} describe the use of OC1 for classifying
2MASS extended sources.  \citet{Bazell01} compared ensembles of
decision trees with ensembles of other types of classifiers.  They
found the largest reduction in classification error was for ensembles
of decision trees.  More recently, \citet{Suchkov05} used an ensemble
of decision tree classifiers to determine types and redshifts of
objects in the SDSS photometric catalog.

The Kohonen Self-Organizing Map (SOM) \citep{Kohonen01} has been used
as an unsupervised method of exploring associations within astronomical data sets
\citep{Mahonen95, Miller96, Naim97a, Rajaniemi02}.  This approach,
an unsupervised neural network, clusters similar
objects together in a way that conserves the topology of the input
data.  In other words, input vectors that are similar to each other
are mapped to neighboring regions of what is usually a two-dimensional
output lattice.  In this way SOMs have been found useful as a way to
visualize high dimensional data.

A common approach to unsupervised clustering is the use of mixture
models \citep{Duda, Mclachlan, Raftery}. Mixture models produce
probabilistic, or soft, assignments of data points to each of the
mixture components, or clusters.  The number of clusters to be learned
must be specified as part of the algorithm.  There is no single
standard approach for choosing this parameter.  Furthermore, without
supervising examples, there is no guarantee that the learned clusters 
will acceptably characterize the known classes within the
data set.

To overcome some of these shortcomings {\it semisupervised learning}
algorithms were proposed, including the work of \citet{Shashahani},
\citet{Miller97}, and \citet{Nigam}.  In these approaches, the input
data consists of both labeled and unlabeled samples.  This affects the
clustering process in two ways.  First the labeled data helps guide
the cluster definitions to properly reflect the known ground-truth
classes \citep{Miller97, Basu1}.  Second, the unlabeled data may help
to more accurately learn parameters of the model, so as to better
define the shapes of the learned classes in feature space
\citep{Shashahani, Miller97}.

Until recently, most semisupervised methods focused on building models
and classifiers for the set of known classes, i.e. those for which
labeled training examples are available.  However, given a data set
with many unlabeled examples, it is quite possible that, latent in the
unlabeled data, some {\it unknown} classes are present.  These unknown
classes are compact clusters of unlabeled objects that have not, as
yet, been recognized by domain experts as distinct classes or
categories of interest.  Such clusters, once identified, could be the
focus of further scientific inquiry, to either validate them as new
classes of interest or to reject them as uninteresting outlier groups.
Recent work in \citet{Miller03a, Miller03b} developed semisupervised
mixture modeling methods with a built-in capability to discover these
unknown classes in mixed labeled and unlabeled data.  Standard
supervised learning algorithms do not have good means to identify
unknown classes because they can only be trained using examples from
the known classes.  Standard unsupervised clustering algorithms make
no distinction between clusters of purely unlabeled data and clusters
containing some labeled samples.  Purely unlabeled clusters are of
particular scientific interest because they represent groups of
objects that are well separated from objects belonging to known
classes.  Thus, they may represent new object types or groups of unusual
objects, i.e, outliers from existing known classes.

In a previous study, \cite{Bazell05} applied the semisupervised class
discovery approach to investigate astronomical data.  In this paper,
we analyze a portion of the SDSS photometric catalog consisting of
approximately 10,000 objects.  The SDSS data in our sample have been
classified via the spectral pipeline into several different types of
objects.  Here we particularly focus on the spectroscopically
unclassified objects, SUOs, which consist of objects that were not
readily classified as stars, galaxies, QSOs, or late type stars (red
stars of type M or later).  The papers describing the SDSS data
releases, in particular, \citet{Stoughton}, as well as the SDSS
website \footnote{http://www.sdss.org} describe the procedure that
leads to objects being labeled as spectroscopically unclassified (the
SDSS class called ``unknown'').  SUOs are objects that passed through
the various filters in the spectroscopic pipeline and were then
examined by a person who could not reliably classify them because they
were too noisy, too featureless, or for some other reason.  We are
interested in applying our semisupervised class discovery technique to
the SDSS data for two main reasons: 1) to see whether this approach
determines significant substructure
within the class of SUOs, i.e., subclasses of SUOs; 2) 
to determine whether some SUOs are clustered into groups that mainly 
consist of known class objects.

An alternative approach would be to apply a simple clustering
algorithm, such as K-means, to identify substructure solely using as
data the set of SUO samples.  However, both the validity of the learned SUO
clusters and the nature of these clusters are best assessed
relative to the known class clusters, by learning a clustering
solution using the data from all the classes, both known and
spectroscopically unclassified.  Our method clusters on the basis of
{\it all} available information -- feature vectors, available class
labels, and the {\it fact} of label presence/absence for each sample
\citep{Miller03a}.  Absence of labels in a compact cluster is suggestive 
that a meaningful subclass of the unknown class may have been found.
\citet{Miller03a} demonstrated that use of label presence/absence
information may help to achieve more accurate clusters and to better
discern unknown subclasses than methods which do not use this
information.

Our semisupervised mixture modeling approach has several benefits not
seen by using standard unsupervised clustering.  In particular, some
objects labeled SUOs may have measured features that are similar to
those of objects from known classes.  In performing clustering using
all the data, both from the known classes as well as the SUO class, we
can learn clustering solutions that reveal these similarities.  That
is, if an SUO is assigned to a cluster of predominantly ``known''
objects of a given type, the clustering is indicating the possibility
that the SUO may be related to (e.g. may be a variant of) the known
object class.  This may also suggest the SUO was mislabeled.
Likewise, consider a cluster that primarily contains SUOs, but which
also owns some known class objects.  The known class composition of
the cluster may hint at the underlying physical nature of the SUO
subclass represented by this cluster.  On the other hand, the
ownership of some known class objects by the cluster may indicate that
these known class objects were mislabeled.  Finally, consider the
problem of choosing the {\it number} of clusters to represent each
class.  Model order selection methods ``match'' the model complexity
(number of clusters in each class) to the amount of data these
clusters own.  If a significant number of SUO class objects can be
explained by, i.e. belong to, clusters of known class objects, then
fewer SUO clusters will be needed in the model.  This means that
learning and model selection using only the SUO class objects could
lead to an overestimate of the number of SUO clusters (subclasses).

There has been limited work using mixture models to classify
astronomical data.  \citet{Strateva01} used a mixture model to verify
the bimodal galaxy distribution they found by other means.
\citet{Nichol01} discuss the use of mixture models for finding
clusters of galaxies.  \citet{Yip04} used a simple mixture model to
identify classes of objects following principal component analysis
(PCA) of spectra.  They applied a mixture model to the histogram of
their parameter $\phi_{KL}$, the angle between the first two
eigencoefficients of the PCA decomposition of the galaxy spectra.
They performed model selection using the Akaike Information Criterion
\citep{Akaike} and found three components best fit their data.
\citet{Kelly04, Kelly05} used a similar approach following PCA of the
shapelet decomposition of galaxy images and used the Bayesian
Information Criterion (BIC, \citet{Schwarz}).  They also found that a
three class representation of their data was optimal, although their
data set was completely different from that of \citet{Yip04}.

\section{Data Preparation}
The Sloan Digital Sky Survey, Data Release 4 contains spectra of about
850,000 objects, categorized into several different classes.  We
selected data with high signal to noise ($SNR > 3.0$) and redshifts
below 2.3.  Because we were especially interested in the class of
SUOs, as determined by the SDSS spectroscopic pipeline, we first
selected objects that met these criteria from the SUO class (SDSS
unknown class).  This resulted in 1763 SUOs.  We then selected 2000
objects each from four other SDSS spectroscopic classes: star, galaxy,
QSO, and late-type star.  For each of these objects we extracted from
the SDSS database the following features: dereddened photometry in
each of the five SDSS bands, $u$, $g$, $r$, $i$, and $z$; four colors,
$(u-g)$, $(g-r)$, $(r-i)$, $(i-z)$; the spectroscopic redshift,
$z_{spec}$, and the spectroscopic class.  This produced our working
dataset of 9763 objects.  The purpose in limiting the study to 10,000
objects, with about 2000 objects from each class, was twofold.  First
we wanted to maintain an approximately equal number of objects from
each class so that the statistics would not be dominated by one class.
In our previous work \citep{Bazell05} we had one class containing
approximately 78\% of the data and several classes with less than 1\%.
Second, we wanted to have a data set that would not take too long to
process.  Computational scaling properties of the algorithm are
discussed in the Appendix.

When running our algorithm we used the four colors as features,
producing a modest 4-dimensional feature space.  While we could also
use the photometric values directly we wanted to be able to compare
our results with those of other groups, for example \citet{Strateva01}
who identified the bimodal galaxy distribution based on color alone.
Each of the objects in our dataset was also associated with the unique
SDSS object identifier which allows us to examine any specific object
in more detail.  The query we used to retrieve these data was:

\begin{verbatim}
Select top 2000
p.objID,
p.dered_u,
p.dered_g,
p.dered_r,
p.dered_i,
p.dered_z,
p.dered_u - p.dered_g,
p.dered_g - p.dered_r,
p.dered_r - p.dered_i,
p.dered_i - p.dered_z,
s.z,
s.specClass
>From BESTDR4..PhotoObj p, BESTDR4..SpecObj s
Where s.bestObjID = p.objID
and s.SpecClass = 0             -- SUO
and s.sn_1 > 3.0
and s.z < 2.3
\end{verbatim}

This SQL command was repeated, changing SpecClass from 0 (SUO) to
1 (star), 2 (galaxy), 3 (QSO), and 6 (late-type star).  The separate
files containing the object IDs and their features were concatenated
and randomized to produce the input file for our algorithm.

\section{Data Modeling Description}
To apply the mixture modeling algorithm we start with a data set where
each object is described by a feature vector. The elements, or
dimensions, of the feature vector represent some measured or derived
quantities from each object.  In the present study the feature vectors
consist of the photometric colors $(u-g)$, $(g-r)$, $(r-i)$, $(i-z)$,
resulting in a 4-dimensional feature space.

The feature vectors are not the only quantities treated as data by our
model.  The other data modeled by our mixture represents the {\it
labeling} information for each object in the data set.  We now discuss
data labeling in more detail.  There are five class types represented
in our data set, the known class categories (star, galaxy, QSO,
late-type star) and the SUO class category.  Each of the 9763 objects
in our set does in fact come with a class label and all these object
labels will be used to evaluate the mixture models that we learn.
However, in order to capture the realistic scenario in which the data
are only {\it sparsely} labeled\footnote{Labeling is in general an
expensive process as it requires substantial human expertise and
hands-on labor.  Thus, in general, even if data samples are abundant,
labeled data may be scarce.}, we treat only a 10\% random sample of
the data from the known classes as labeled for purposes of model
learning/clustering, with all the other known class objects treated as
unlabeled.  While we used a single 10\% random sampling of the data,
we performed several runs using this data sample with different
random initial conditions for the algorithm.  We discuss this in more
detail in the Results section.

The SUO class is one of the class types.  Thus, all the SUOs are
in fact labeled.  However, there is uncertainty about the nature
of each of the objects in this class.  Some SUOs may represent noisy
measurements or objects of little interest, resulting from bad columns
on the CCD or part of the spectrum falling off the edge of the CCD.
Some may be objects from known categories that were not properly
labeled as such.  Some may represent new object types of
scientific interest.  To reflect our genuine uncertainty about the
nature of these spectroscopically unclassified objects, we treat them
all as {\it unlabeled}.  The goal of our mixture modeling is to try to
discern the underlying structure in and nature(s) of these objects.

Consistent with this discussion, for purposes of data modeling, each
labeled object is described by its feature vector, its class label,
and a symbolic value indicating that the label is {\it present} for
the object.  Each unlabeled object is described by its feature vector
and a symbolic value indicating that the label is {\it absent}.  The
model is required to explain all the data describing each
object, including the presence or absence of the class label.
Necessitating the explanation of this labeling information encourages
the model to learn mixture solutions with some clusters that represent
{\it purely unlabeled} (or nearly so) object subsets, with
other clusters containing a mix of labeled and unlabeled objects -- while the
latter clusters represent known class data, the former
may represent unknown subclasses.

To facilitate the learning of these mixture solutions, we define two
different types of mixture components we dub ``predefined'' and
``nonpredefined''.  Predefined components generate data points that
can be either labeled or unlabeled, but where the labels are assumed
to be {\it missing at random}.  These components represent the known
classes (for which class labels should, in fact, be missing at random,
consistent with a random sample of the data being
labeled). Nonpredefined components generate only unlabeled data
points, i.e., the labels are always missing from the data.  One way to
picture this is to imagine a cluster of data points in some feature
space.  If that cluster of points contains a random mixture of labeled
and unlabeled points it will be described by a predefined component.
If the cluster contains essentially all unlabeled points it will be
described by a nonpredefined component \footnote{The definition of
``essentially all'' is determined by an adjustable threshold parameter
in our model.  This parameter can be set either manually or
automatically \citep{Miller03b, Bazell05}.  For purposes of this
paper, if greater than 95\% of the objects described by a given
component are unlabeled, then the component is declared
nonpredefined.}. The nonpredefined components contain 
data points that are almost purely unlabeled, and may represent
novel classes or subclasses of objects.

The mixture model is learned based on local maximization of a
statistical likelihood function.  The ``best fit'' parameters that describe the
mixture model, including Gaussian parameters (means and variances),
the coefficients of the mixture components (which are prior
probabilities) and the distribution used to model label generation
given a particular mixture component, are determined by locally
maximizing this function.  This produces a set of probabilities that
each mixture component describes a given class, with ``unlabeled''
treated as an additional class value.  If the probability is high
(close to one) that a specific mixture component describes unlabeled
data, then that component is declared nonpredefined and,
putatively,
this component may be describing a new class.  We provide
further details of the algorithm in the Appendix and in
\citet{Miller03a, Miller03b, Bazell05}.

As mentioned, one of the parameters learned by the mixture model is
the variance of each Gaussian component.  If the variance is small
then the component is compact and spans only a small part of parameter
space.  This tends to increase the number of components needed to
describe the data, increasing the model complexity.  Although we use
BIC to determine the optimal model order, our previous work
\citep{Bazell05} showed that BIC sometimes had trouble converging on a
finite model order.  To avoid this we applied a variance threshold,
essentially a minimum allowed width of the Gaussian component.  This
alleviated the convergence problem and resulted in simpler models.  We
use the same variance thresholding technique in this work.

For each run of the algorithm we must specify the number of mixture
components to be used -- the model order.  Determining the correct
model order can be difficult and there are several proposed criteria
in the literature \citep{Schwarz, Wallace, Mclachlan}.  However, there
is no agreed upon method appropriate to a given situation.  Model
order selection is especially important in the case of class discovery
because we view the nonpredefined components in the model as potential
new classes. Furthermore, the distribution of objects among the
predefined components clearly changes with the model order.  Accurate
model order selection is thus important for classification and
successful new class discovery.  Here, as in our previous work
\citep{Miller03a, Miller03b, Bazell05} we run the algorithm for a
range of model orders and use a procedure called the Bayesian
Information Criterion (BIC) \citep{Schwarz} to choose the best one.
See the Appendix for more details on BIC.

We ran the semisupervised discovery code using from 5 to 70 components
and allowed BIC to choose the best model order.  After our initial run
we found that the minimum BIC cost was for a model with 16 components
given the four features we used, although the minimum was quite broad.
We then ran the code five more times, with different random
initializations of the parameters, bracketing the initial minimum
between 10 and 30 components.  The minimum BIC cost remained at a 16
component model. We used this model for subsequent investigations.

We also compared the semisupervised model with a completely
unsupervised mixture model.  This allowed a useful comparison of the
techniques and a fuller understanding of the benefits of the
semisupervised approach.

It is worth pointing out that changing the number of features and data 
points will in
general change the optimal number of components that are found.
For example, in our previous work \citep{Bazell05} we used
six features (and 50,000 objects) and found optimal model orders in
the range of 60-70 components.

\section{Results} 
Figure 1 shows the BIC cost for our data as a function of the number
of components for all the runs performed. The minimum is broad and the
BIC cost varies significantly over the repeated trials between 10 and
30 components.  The global minimum at 16 components is evident,
although the 14 and 24 component models do almost as well.  These runs
were for a single 10\% random sampling of the data.  Ideally we would
have several independent random samples and run multiple trials and a
range of model orders.  However, this becomes computationally
expensive so we chose a single random sample of the data with a modest
number of trials over a modest range of model orders.

\clearpage

\begin{figure}
\epsscale{0.8}
\plotone{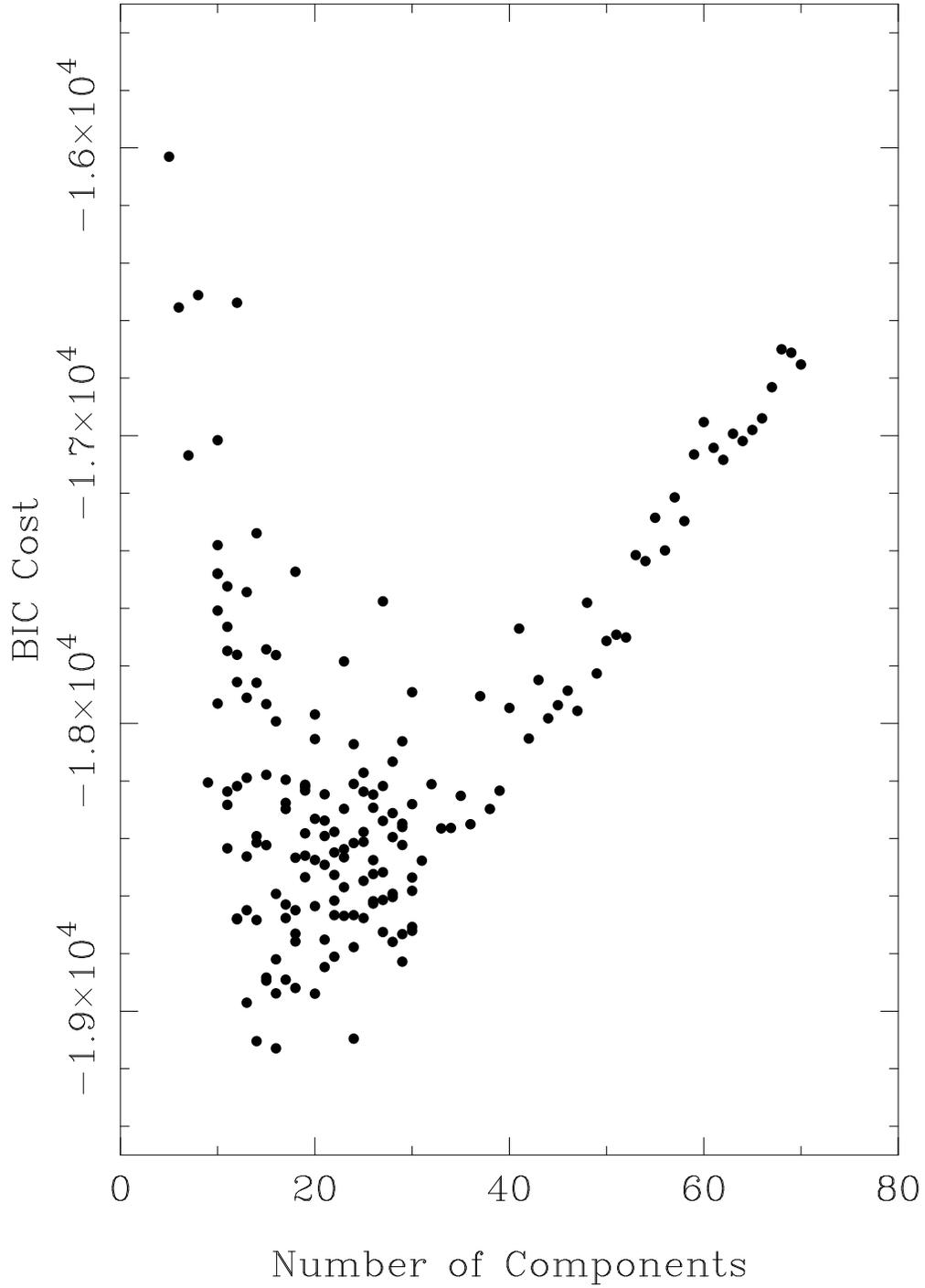}
\label{fig:BIC}
\caption{\small BIC cost as a function of number of components in the
  mixture model.  The minimum cost is for 16 components with 14 and 24
  component models having a slightly higher cost.}
\end{figure}

\clearpage

Of the 16 components in the best model, 14 were predefined and 2 were
nonpredefined.  While the mixture model assigns probabilities of class
membership to each point in the data set (``soft'' assignments) we
chose the class with the highest probability and assigned each point
to that class (``hard'' assignments).  Thus, some points may have
significant probabilities of membership in more than one class.
Conversely, each component will own data points from several different
classes.  Some components will be more pure than others.  Table
\ref{tab:comp-membership} shows the number of data points from each
class that are owned by each component in the model (hard assignment).
Components 10 and 11 are nonpredefined components, meaning that the
clusters of points owned by these components are almost purely
unlabeled.  As we can see from Table \ref{tab:comp-membership} both of
these components have a majority of their points from the SUO
class (0), but component 10 contains a significant numbers of points
from the star class.
\clearpage
\begin{deluxetable}{rrrrrrr}
\tabletypesize{\scriptsize} 
\tablecaption{Number of objects from each class assigned to each
  component:  semisupervised algorithm.
  \label{tab:comp-membership}}
\tablewidth{0pt}
\tablehead{
\colhead{Comp} & \colhead{np} & \colhead{SUOs} &
\colhead{Stars} & \colhead{Galaxies} & \colhead{QSOs} &
\colhead{Late-Type Stars}}
\startdata
0 & 0 &  0 &   0&    0&    0&  494 \\
1 & 0 & 66 &  17&  147&    1&  114 \\
2 & 0 & 16 &   5&   39&    0&  650 \\
3 & 0 & 99 &  31&    1&    8&    3 \\
4 & 0 &  9 &   2&    2&    0&  132 \\
5 & 0 & 60 &  40&    3&    3&  125 \\
6 & 0 & 10 & 769&    9&    2&    0 \\
7 & 0 & 17 &  11&    0&    0&   58 \\
8 & 0 & 70 & 469&  167&   25&    1 \\
9 & 0 & 152&  17&    2&  221&   10 \\
10& 1 & 360& 250&    2&    5&    1 \\
11& 1 & 296&   9&    2&    0&   19 \\
12& 0 & 325&  48&   12& 1599&    0 \\
13& 0 &  12&  47&    2&    0&    0 \\
14& 0 & 109& 179&  960&   15&  345 \\
15& 0 & 162& 106&  652&  121&   48 \\
\enddata
\tablecomments{np = 1 denotes nonpredefined components.}
\end{deluxetable}
\clearpage
Component 0 contains only late type stars while components 2 and 4
contain 92\% and 91\% late type stars, respectively.  Component 12,
with 81\% quasars, has the largest number of quasars, and the largest
total number of objects of any component.  It also owns 325 SUOs.
This component is significantly contaminated with SUOs and, to a
lesser extent, stars.  Component 9 also has a relatively large number
of quasars (221 or 55\%) and a significant number of SUOs also (152 or
38\%).  Component 15 has 121 quasars but is dominated by galaxies.  We
will discuss components 14 and 15 in more detail later.  The
components with large numbers of galaxies are often contaminated with
stars or late type stars (components 1, 8, 14, and 15) and objects of
type SUO are also often present.  Components 14 and 15 together
own 81\% of the galaxies in the sample while component 12 alone owns
80\% of the quasars.

Of the 1763 objects of type SUO, the nonpredefined components 10
and 11 together own 656 or 37\%.  It is clear from Table
\ref{tab:comp-membership} that the SUOs are spread across
the components.  Component 11 consists of 91\% SUOs with
the remaining 9\% objects almost entirely stars and late-type stars.
Only one quasar made it into this nonpredefined component.  The
SUOs are spread widely among the remaining components,
with 6 components (9,10,11,12,14, and 15) needed to account for more
than 75\% of these objects.

Table \ref{tab:phot-stats} shows some additional statistics for each
component, displaying the mean, median, and rms for each component in
each of the four colors we used as features.  Note that component 10,
which has 58\% of its points from the SUO class, has a relatively
small rms deviation from the mean for each of the colors even though
it contains a mixture of different object types.  Conversely,
component 11, which is 91\% SUO class 0, has a much larger
standard deviation for $(u-g)$ in particular.  Note also that component
12, which is mainly quasars but also has a significant number of
SUOs, has quite a small rms in all colors.
\clearpage
\begin{deluxetable}{rrrrrrrrrrrrrrrr}
\tablecolumns{16}
\tabletypesize{\scriptsize}
\tablecaption{Statistics for best model components by BIC
  \label{tab:phot-stats}}
\tablewidth{0pt}
\tablehead{
\colhead{} & \multicolumn{3}{c}{$(u-g)$} & 
\colhead{} & \multicolumn{3}{c}{$(g-r)$} & 
\colhead{} & \multicolumn{3}{c}{$(r-i)$} &
\colhead{} & \multicolumn{3}{c}{$(i-z)$}\\
\cline{2-4} \cline{6-8} \cline{10-12} \cline{14-16}\\
\colhead{Component} & 
\colhead{Mean} & \colhead{Median} & \colhead{rms} & \colhead{} &
\colhead{Mean} & \colhead{Median} & \colhead{rms} & \colhead{} & 
\colhead{Mean} & \colhead{Median} & \colhead{rms} & \colhead{} & 
\colhead{Mean} & \colhead{Median} & \colhead{rms}}
\startdata
0 & 2.43 & 2.38 & 1.01 & & 1.56 & 1.55 & 0.12 & & 1.98 & 1.95 & 0.17 & 
& 1.08 & 1.06 & 0.10 \\
1 & 2.02 & 2.11 & 0.75 & & 1.73 & 1.72 & 0.26 & & 0.68 & 0.67 & 0.14 & 
& 0.38 & 0.39 & 0.10\\
2 & 2.82 & 2.69 & 0.90 & & 1.49 & 1.45 & 0.17 & & 1.00 & 1.00 & 0.25 & 
& 0.56 & 0.56 & 0.13 \\
3 & 1.75 & 1.90 & 0.68 & & 0.81 & 0.77 & 0.26 & & 0.29 & 0.29 & 0.14 & 
& 0.07 & 0.00 & 0.30\\ 
4 & 1.55 & 1.57 & 0.93 & & 1.60 & 1.63 & 0.56 & & 1.09 & 1.16 & 0.40 & 
& 0.72 & 0.73 & 0.18 \\
5 & 1.77 & 1.83 & 1.17 & & 0.79 & 0.84 & 0.81 & & 0.81 & 0.76 & 1.03 & 
& 0.40 & 0.45 & 0.95 \\
6 & 1.04 & 1.04 & 0.15 & & 0.07 & 0.09 & 0.15 & & -0.01 & -0.00 & 0.10 & 
&-0.05 & -0.04 & 0.08 \\ 
7 & 0.85 & 0.80 & 2.96 & & 2.47 & 1.94 & 2.79 & & -0.88 & -0.53 & 2.96 & 
& 1.44 & 0.82 & 2.91 \\
8 & 1.06 & 1.06 & 0.20 & & 0.39 & 0.38 & 0.10 & & 0.18 & 0.16 & 0.11 & 
& 0.06 & 0.05 & 0.07 \\ 
9 & 0.38 & 0.36 & 2.96 & & 0.33 & 0.33 & 0.16 & & 0.24 & 0.26 & 0.16 & 
& 0.30 & 0.28 & 0.14 \\
10 &0.07 & 0.10 & 0.22 & & -0.23 & -0.19 & 0.15 & & -0.21 & -0.20 &
-0.10 & & -0.26 & -0.25 & -0.16 \\
11 &3.43 & 3.40 & 0.70 & & 0.87 & 0.83 & 0.23 & & 0.31 & 0.29 & 0.13 & 
& 0.09 & 0.13 & 0.21 \\
12 &0.19 & 0.18 & 0.16 & & 0.10 & 0.09 & 0.13 & & 0.09 & 0.08 & 0.14 & 
& 0.02 & 0.01 & 0.12 \\
13 &1.86 & 1.84 & 0.24 & & 0.66 & 0.69 & 0.15 & & 0.24 & 0.24 & 0.11 & 
& 0.16 & 0.14 & 0.09 \\
14 &2.04 & 1.96 & 0.35 & & 1.07 & 1.03 & 0.19 & & 0.46 & 0.45 & 0.12 & 
 & 0.32 & 0.33 & 0.09 \\
15 &1.24 & 1.30 & 0.36 & & 0.67 & 0.67 & 0.13 & & 0.36 & 0.37 & 0.12 & 
& 0.26 & 0.25 & 0.11 \\
\enddata
\end{deluxetable}
\clearpage
In Figure 2 we show three color-color diagrams for our entire data
set.  Points plotted as green are from nonpredefined components 10 and
11, while the blue points are from the other components.  The contour
levels are at 0.8, 0.6, 0.4, 0.2, 0.1 and 0.05 of the maximum and are
plotted on a linear scale.  It is evident from the color-color
diagrams that the points from these two components are well separated
and do not overlap with each other.  We emphasize that these two
components define regions in the feature space containing almost
entirely unlabeled data points.  This figure shows projections of the
4-dimensional feature space onto three 2-dimensional planes.  While
points from these components appear to be overlapping the blue points
in projection, they are actually well separated in the 4-dimensional
feature space.  Each of these components is shown in more detail in
Figure 3 which shows only the points from components 10 and 11.  The
points are color and numerically coded for the spectroscopic class
they were assigned: green 0--SUO; blue 1--star; orange 2--galaxy;
red 3--quasar; cyan 4--late-type star.  The $(u-g)$ vs $(g-r)$ diagram shows
that component 10 has the SUOs clustering strongly between
$-0.1<(u-g)<0.3$ and $-0.3<(g-r)<0$ although there is a faint tail of
bluer objects.  Similar clustering is evident in the $(g-r)$ vs. $(r-i)$
diagram.  However, the $(r-i)$ vs. $(i-z)$ diagram appears to show the SUO
class objects to be more spread out and intermixed with the other
classes.

\clearpage

\begin{figure}
\epsscale{0.8}
\plotone{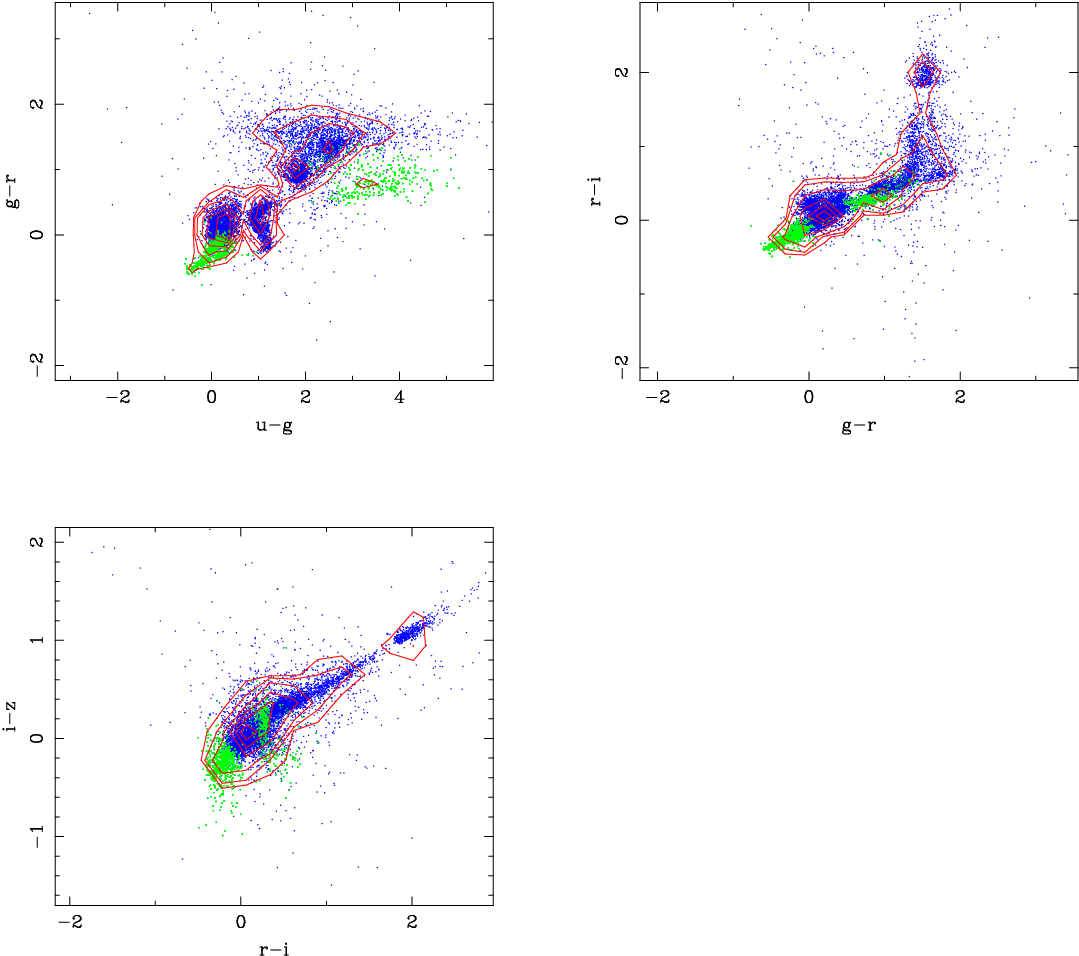}
\label{fig:color-color}
\caption{\small Color-Color diagrams for all data used.  Green dots
  represent data from components 10 and 11.  Blue dots represent data
  from all other components.  Component 10 is the bluer cluster of
  points (down and to the left); component 11 is the redder cluster.
  Contours are for all the data with levels 5\%, 10\%, 20\%, 40\%,
  60\% and 80\% of the maximum.}
\end{figure}

\begin{figure}
\epsscale{0.8}
\plotone{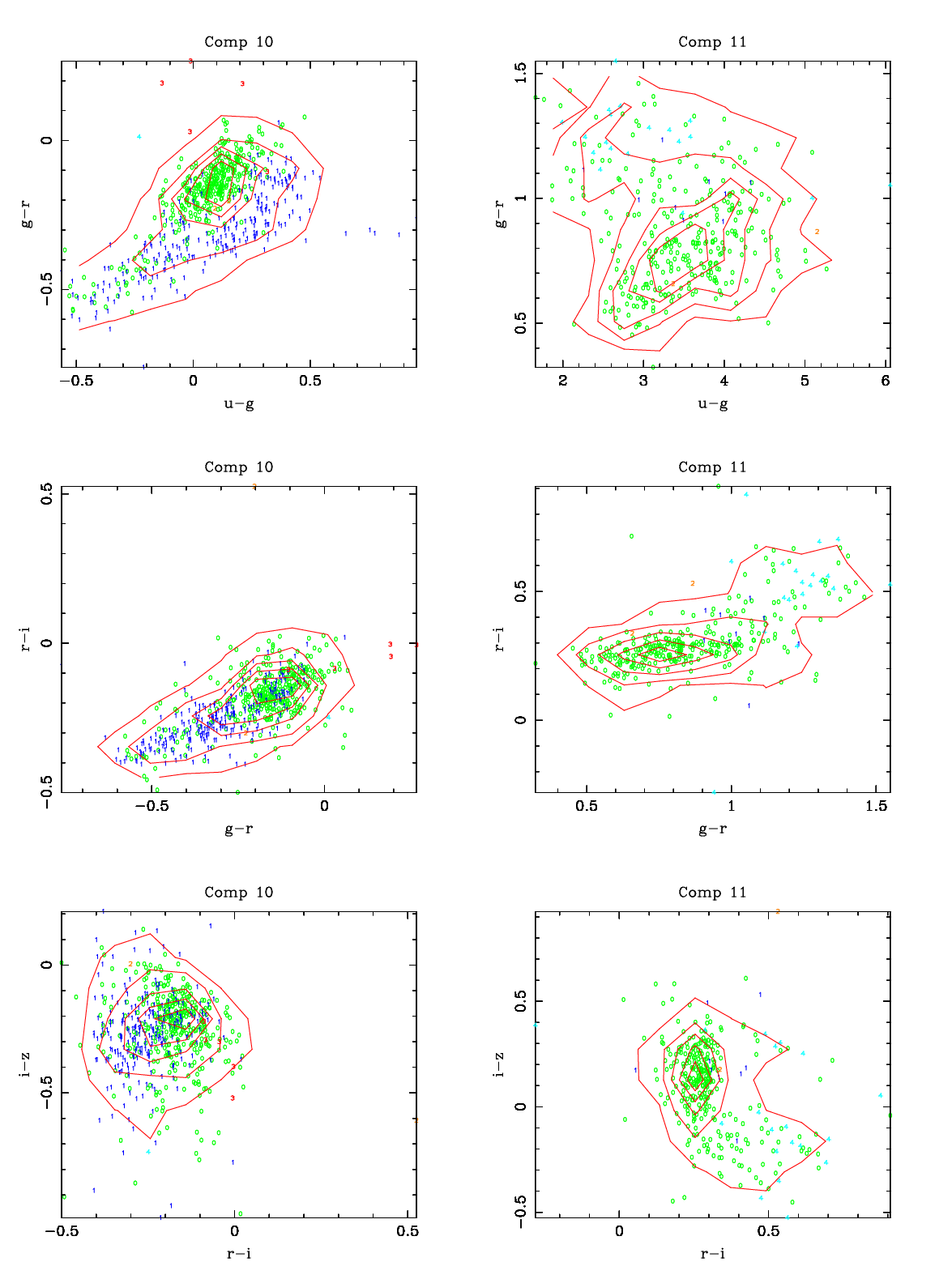}
\label{fig:color-color10-11}
\caption{\small Color-Color diagrams for components 10 and 11.  The
   symbols (colors refer to the online version) are: green 0--SUO,
   blue 1--star, orange 2--galaxy, red 3--qso, cyan 4--late type star.
   Component 10 has the SUOs tightly clustered and fairly
   well separated from the stars.  This part of color-color space
   overlaps regions containing stars and quasars.  Component 11
   contains 80\% SUO and is much redder than component 10.
   Contours are for all the data with levels 5\%, 10\%, 20\%, 40\%,
   60\% and 80\% of the maximum.}
\end{figure}
\clearpage

We also ran an unsupervised mixture model on the same data set in
order to compare the results of the semisupervised approach with the
unsupervised approach.  Again, we ran five trials of the unsupervised
mixture model and found the best solution using the Bayesian
Information Criterion.  These tests resulted in an optimal mixture
model consisting of 15 components.  Table \ref{tab:comp-membership2}
summarizes the number of objects from each class that were assigned to
each of the 15 components.  Comparison of the data in this table with
the data from table \ref{tab:comp-membership} for the semisupervised
case provides a qualitative understanding of the differences between
the two methods.

\clearpage
\begin{deluxetable}{rrrrrr}
\tabletypesize{\scriptsize} 
\tablecaption{Number of objects from each class assigned to each
  component: unsupervised clustering.
  \label{tab:comp-membership2}}
\tablewidth{0pt}
\tablehead{
\colhead{Comp} & \colhead{SUOs} &
\colhead{Stars} & \colhead{Galaxies} & \colhead{QSOs} &
\colhead{Late-Type Stars}}
\startdata
0 &  66 &  17&  139&    1&  184 \\
1 &   0 &   0&   0 &    0&  487 \\
2 & 208 &  29&   19& 1488&    3 \\
3 &  94 &  44&    7&    4&  190 \\
4 & 320 &  17&    2&    0&    7 \\
5 & 170 & 210& 1261&   31&   87 \\
6 & 234 &  46&    1&  262&    0 \\
7 &  17 &  13&    0&    0&   64 \\
8 & 110 &  11&    6&  139&   32 \\
9 &   3 &   2&    1&    0&  446 \\
10&  52 &  61&  148&    4&  492 \\
11&  89 &  95&  291&   53&    6 \\
12& 315 & 235&    1&    4&    1 \\
13&  80 & 587&  121&   14&    1 \\
14&   5 & 633&    3&    0&    1 \\
\enddata
\end{deluxetable}
\clearpage

\section{Discussion}
The plot of the BIC cost vs. number of components, Figure 1, shows a
distinct global minimum, but there is also a lot of scatter evident.
We performed additional runs using models with between 10 and 30
components in an attempt to quantify the reproducibility of the number
of nonpredefined components in each model. Table \ref{tab:multirun}
displays some results that illustrate this issue.

\begin{deluxetable}{rrrrr}
\tabletypesize{\scriptsize} 
\tablecaption{Number of Nonpredefined Components Resulting from Repeated
  Algorithm Runs
  \label{tab:multirun}}
\tablewidth{0pt}
\tablehead{
\colhead{$N_{comp}$} & \colhead{$N_{np}$} & \colhead{Avg} &
\colhead{$\sigma$}}
\startdata
10 & 1 2 2 3 2 & 2.0 & 0.63 \\
11 & 2 2 1 2 2 & 1.8 & 0.40 \\
12 & 2 2 2 2 3 & 2.0 & 0.40 \\
13 & 2 1 2 2 3 & 2.0 & 0.63 \\
14 & 2 3 1 2 2 & 2.0 & 0.63 \\
15 & 2 3 2 1 5 & 2.8 & 1.47 \\
16 & 3 3 3 4 2 & 3.0 & 0.63 \\
17 & 3 4 3 2 3 & 3.0 & 1.26 \\
18 & 4 1 2 4 4 & 3.0 & 1.26 \\
19 & 5 2 4 5 6 & 4.4 & 1.46 \\
20 & 3 2 6 6 3 & 4.0 & 1.67 \\
21 & 5 3 4 3 3 & 3.6 & 0.80 \\
22 & 3 3 5 4 3 & 3.6 & 0.80 \\
23 & 3 6 5 2 4 & 4.0 & 1.41 \\
24 & 8 7 4 6 4 & 3.4 & 1.60 \\
25 & 3 4 5 2 3 & 3.4 & 1.02 \\
26 & 5 3 4 4 2 & 3.6 & 1.02 \\
27 & 7 4 3 3 6 & 4.6 & 1.62 \\
28 & 5 5 5 5 4 & 4.8 & 0.40 \\
29 & 6 6 5 2 4 & 4.6 & 1.50 \\
30 & 6 4 5 5 7 & 5.4 & 1.02 \\
\enddata
\tablecomments{Column $N_{np}$ denotes the number of nonpredefined
  components for each of 5 runs with a total of $N_{comp}$ components.}
\end{deluxetable}

Column 1 gives the number of components in the model.  Column 2 gives
the number of nonpredefined components for each of the five additional
runs performed.  Column 3 gives the average number of nonpredefined
components and Column 4 gives the standard deviation.  We see that
there is a trend toward increasing number of nonpredefined components
as the model complexity (total number of components) increases.  This
is expected since {\it all} classes, including the SUO class, should
be represented by more components as the total number of components is
increased.

Note that for $N_{comp} = 16$ the additional model runs produced an
average of 3 nonpredefined components.  However, these models all had
higher BIC cost than the one we chose for analysis, which was based on
the model with the lowest BIC cost; this model had only two
nonpredefined components.  For $N_{comp} = 14$, which also had a low
BIC cost, the average number of nonpredefined components is 2, with
little scatter.  For $N_{comp} = 24$, the other model with especially
low BIC cost, the average number of nonpredefined components is 3.4,
with three models producing six or more nonpredefined components.
This is significantly different from the model we analyzed in detail
and deserves further study.  We hope to examine this in more detail in
a future paper.

The comparison of Table \ref{tab:comp-membership} with Table
\ref{tab:comp-membership2} provides some insight into the different
results that are achieved with the semisupervised vs. unsupervised
algorithms.  The best unsupervised model had 15 components compared
with the best semisupervised model, which contained 16 components.
Note that the numbers identifying each component are arbitrary so we
cannot perform a direct component to component comparison between the
two models.  However, we do note the following.  Component 0 of the
semisupervised model (SS0 for short) is very similar to component 1 of
the unsupervised model (US1 for short); they captured only late-type
stars, with 494 and 487 stars respectively.  Also, SS12 corresponds
closely to US2 since both captured the bulk of the QSOs in the sample
(80\% and 74\% respectively).  Still, the semisupervised algorithm did
a better job at isolating the QSOs.  SS10 and SS11, the nonpredefined
components, appear to have unsupervised analogs in US12 and US4.  SS10
and SS11 together capture 37\% of the SUOs while US12 and US4 capture
36\%.

According to the SDSS spectroscopic classification procedure, the
objects that comprise component 10 are a mixture of mainly SUOs and
stars.  It is notable in Figure 3 that many of the SUO class points
are strongly clustered together between $-0.1<(u-g)<0.3$ and
$-0.3<(g-r)<0$. This component overlaps the white dwarf exclusion
region of \citet{Richards02} in all three color-color projections they
defined.  It also overlaps a region of low redshift, ugri selected
quasars in $(u-g)$ vs. $(g-r)$ space (see figure 13 of
\citet{Richards02}; also \citet{Hall02}). It also overlaps the high
density region of confirmed quasars from \citet{Richards04} in all
three projections.  All the works just cited base their quasar
identifications on photometric properties, while our identification of
the objects as SUOs is based on their spectral properties being
unusual in some way.  Indeed, a visual inspection of the spectra
labeled as SUOs in component 10 shows that they are almost all blue,
relatively featureless spectra.  Most of these are likely to be stars,
sdO or DA white dwarfs \citep{Kleinman04}, with a small number of BL
LAC objects.  Figure 5 shows spectra from several representative
objects in component 10.
\clearpage
\begin{figure}
\epsscale{0.8}
\plotone{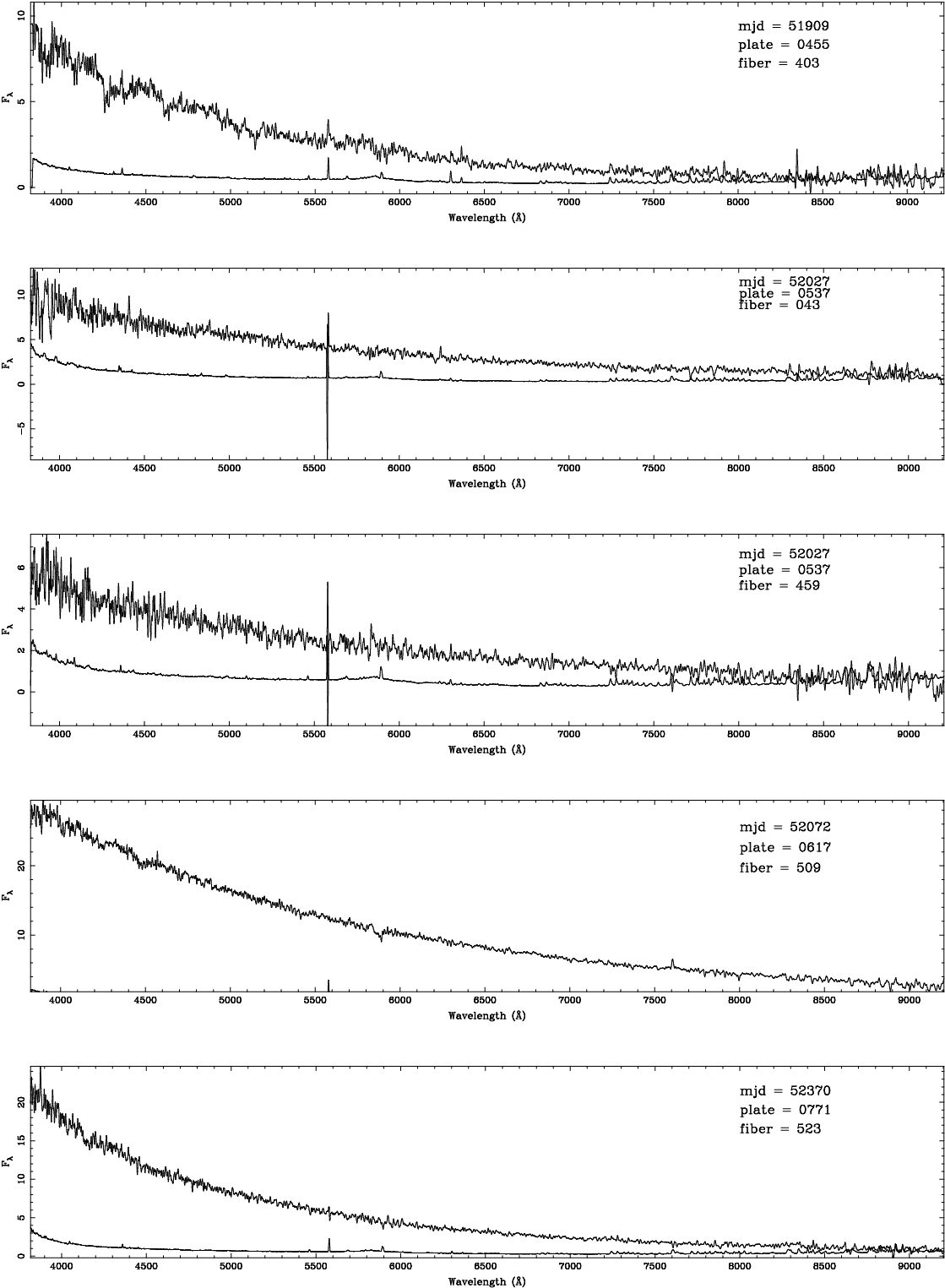}
\label{fig:Spec10}
\caption{\small Typical spectra from objects in component 10.  These
  are all blue, featureless spectra which are most likely sd0 or DA
  white dwarfs.  The lower curve in each plot is the associated error.
  The spectra have been smoothed to about $5\AA$ resolution.}
\end{figure}
\clearpage
Component 11 in our mixture model has very different characteristics
from component 10.  It consists of almost purely SUO type objects
(80\%), and it has a much broader distribution of all four colors.
Furthermore, all the points in component 11 are much redder than those
in component 10.  Component 11 has captured a region of color space
that is largely in the HiZ QSO region of the $(u-g)$ vs. $(g-r)$ color-color
diagram of \citet{Richards02}.  However, it also corresponds to a
region of relatively high density of objects initially classified as
quasars but then rejected following a cut on stellar density (see
figure 2 of \citet{Richards04}).  A visual inspection of component 11
objects labeled as SUOs indicates that a majority of these objects
are low signal to noise G-K stars.  See Figure 6 for sample spectra
from objects in component 11.  SDSS target selection is fainter for
QSOs than stars and galaxies since their prominent broad features make
QSOs easier to identify at low signal to noise.  Stars incorrectly
targeted as QSOs are thus more likely to be classified as SUOs.
\clearpage
\begin{figure}
\epsscale{0.8}
\plotone{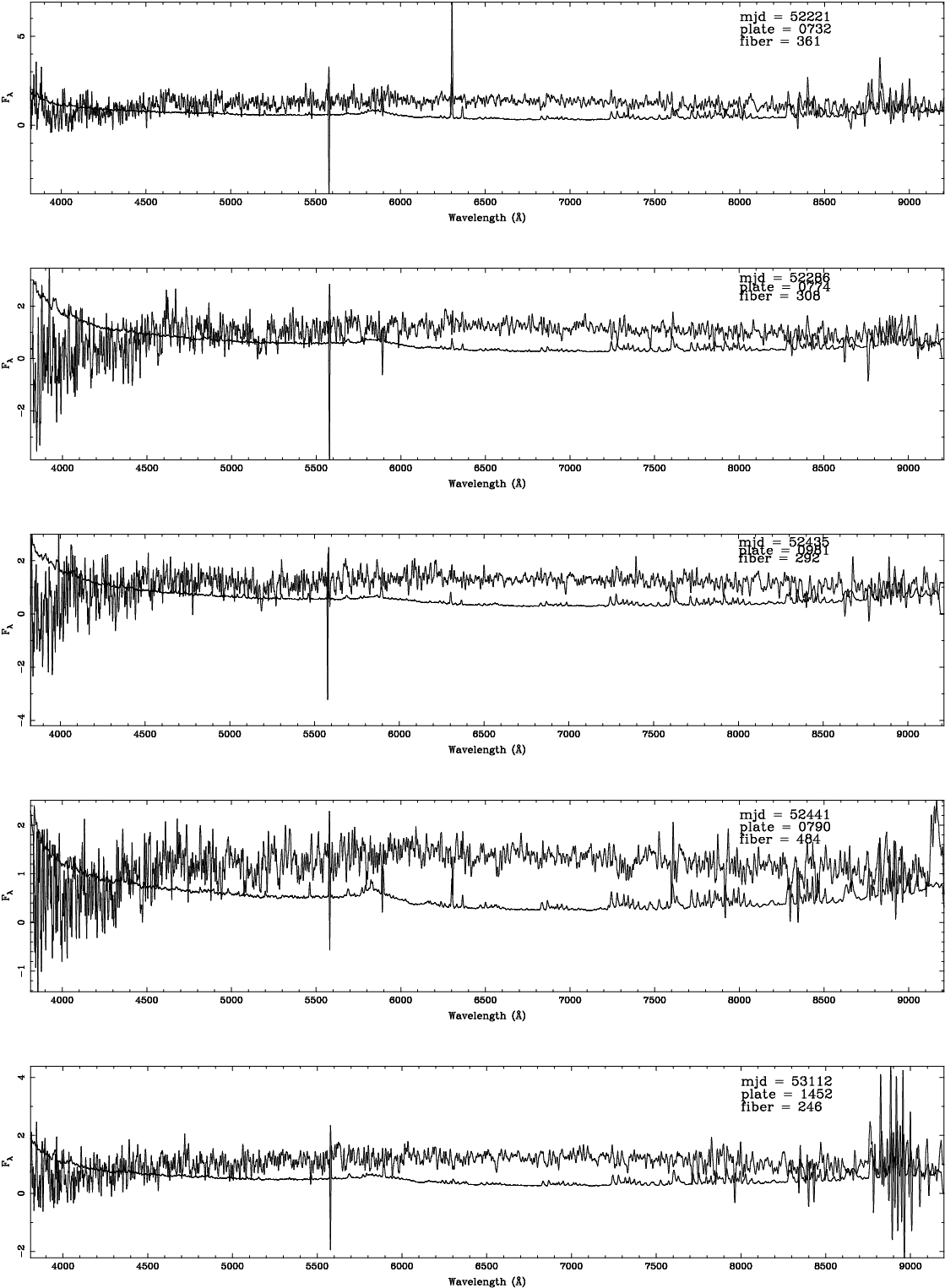}
\label{fig:Spec11}
\caption{\small Five typical spectra of objects from component 11.
  These are probably low SNR G-K stars.  The relatively smooth line is
  the error level.  The spectra have been smoothed to about $5\AA$
  resolution.}
\end{figure}
\clearpage
The small rms values for several of the components suggest, on the
surface, a homogeneity of object types within the components.  For
example, component 10 has a relatively small rms in all colors (see
Table \ref{tab:phot-stats}).  This is consistent with all the objects
being owned by a single mixture component but it does not necessarily
mean all the objects are the same.  That determination would have to
be made with more detailed examination and comparison of the spectra
of the objects.

Moreover, there are several possible reasons why the nonpredefined
components contain significant numbers of points from other classes.
First, for some objects, the fact that the object is assigned to the
SUO class is indicative that there is significant uncertainty about
its class of origin.  All objects that are identified as SUOs by the
SDSS spectral pipeline are also visually inspected.  Some objects in
the SUO class may really belong to one of the known classes, but were
not classified as such in the SDSS spectral pipeline because the
specificity of the pipeline (its ability to identify positive
instances of an object class) is limited.  The procedure for labeling
an object as spectroscopically unclassified involves cross-correlation
with several standard templates and the determination of the
confidence level of the cross-correlation.  When this confidence level
is below 0.25 then the object is labeled as an SUO.  Subsequent visual
inspection was unable to provide a confident classification of these
objects.  If there are a significant number of such objects and if
they have similar feature vectors, a nonpredefined cluster may be
learned which contains many of these objects (and whose model
parameters well-describe these objects).  However, such a cluster may
also well-describe (unlabeled) known class objects, and thus may
contain a significant number of such objects.  Another possibility is
that some objects in the known classes are mislabeled and should
really be members of the SUO class or classes.

While the mixed composition of nonpredefined components is partially
explained by uncertainty and/or errors associated with the spectral
pipeline object labelings, another possible explanation is that the
features we chose were not powerful enough to fully distinguish
between objects from the different classes.  In particular, since each
photometric band is essentially a weighted average of the spectrum of
the object, it is clear that significantly different spectra may
produce similar photometric responses.  This suggests that one should
be cautious when using only photometric features for classification
purposes.  Using additional features, such as the photometry in $u$,
$g$, $r$, $i$, and $z$, might help differentiate between
classes. Certainly, including UV or IR data in a multispectral
analysis would lead to more powerful discriminators.  Working directly
with the high-dimensional object spectra would also substantially
enhance the potential for class discrimination.  However, there are
also many spectral features that are {\it not} class-discriminating.
This indicates the need for effective feature selection, to determine
the (perhaps small) subset of features that are most important for
distinguishing the different object classes.  Some recent approaches
have been proposed for feature selection in high-dimensional mixture
modeling, e.g. \citep{Graham06}, which we hope to exploit in the near
future.

\clearpage

\begin{figure}
\epsscale{0.8}
\plotone{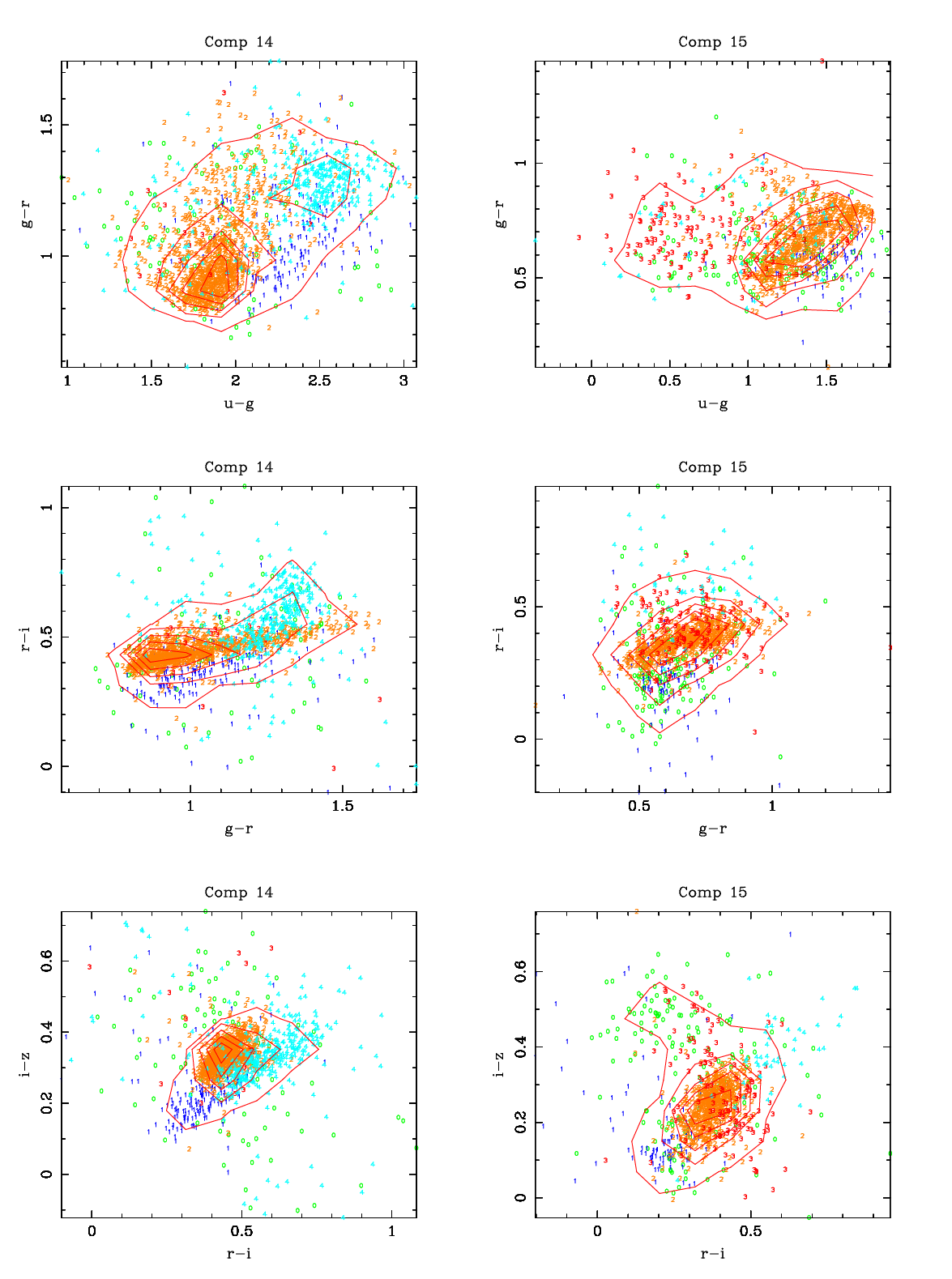}
\label{fig:color-color14-15}
\caption{\small Color-Color diagrams for components 14 and 15.  The
  symbols (colors refer to the online version) are: green 0--SUO,
  blue 1--star, orange 2--galaxy, red 3--qso, cyan 4--late type star.
  Component 15 contains mainly bluer objects, of which 56\% were
  spectroscopically identified as galaxies.  Component 14 contains
  redder objects, of which 58\% are identified as galaxies Contours
  are for all the data with levels 5\%, 10\%, 20\%, 40\%, 60\% and
  80\% of the maximum.}
\end{figure}

\clearpage

As mentioned above, a number of predefined components also have
significant admixtures of various types of objects.  In particular we
note components 14 and 15, which together have 81\% of all the
galaxies in our sample.  These two components split the galaxies into
two clusters, as shown in Figure 6.  This is very similar to the
bimodal distributions found by \citet{Strateva01} in the SDSS data and
more recently seen also in combined GALEX and SDSS observations
\citep{Seibert05}.  The two components contain essentially equal
fractions of galaxies (60\%), though some fraction of the SUO class
objects could also be galaxies.  They also have a low fraction of
quasars.  Our modeling procedure allows for the possibility that
several mixture components may be needed to describe a single class.
However, in this case no other components appear to overlap this
region of color-color space.  It is hard to find components for the
unsupervised model which correspond well to components SS14 and SS15
that capture some of the bimodal galaxy distribution.  US5 alone
captures 63\% of the galaxies, but the remaining galaxies are fairly
evenly spread among four other components.

Notably, the separation into two clusters by the semisupervised
mixture model is a byproduct of our main objective of looking for
subclasses in the SUOs.  However, even with a sample containing only
2000 galaxies, we find that our mixture model does a very good job of
separating the galaxy distribution into two separate clusters.  It
also appears to result in a better separation than that due to the
unsupervised approach.  We hope to perform a similar analysis using a
much larger sample of galaxies and look for a more detailed
decomposition of the main galaxy class.

\section{Acknowledgments}
We would like to thank the NASA Applied Information Systems Research
Program for supporting us in this effort under contract NAS5-02098.
We also extend our thanks to the anonymous referee for a number of
suggestions that greatly improved this paper.

Funding for the creation and distribution of the SDSS Archive has been
provided by the Alfred P. Sloan Foundation, the Participating
Institutions, the National Aeronautics and Space Administration, the
National Science Foundation, the U.S. Department of Energy, the
Japanese Monbukagakusho, and the Max Planck Society. The SDSS Web site
is http://www.sdss.org/.

The SDSS is managed by the Astrophysical Research Consortium (ARC) for
the Participating Institutions. The Participating Institutions are The
University of Chicago, Fermilab, the Institute for Advanced Study, the
Japan Participation Group, The Johns Hopkins University, the Korean
Scientist Group, Los Alamos National Laboratory, the
Max-Planck-Institute for Astronomy (MPIA), the Max-Planck-Institute
for Astrophysics (MPA), New Mexico State University, University of
Pittsburgh, University of Portsmouth, Princeton University, the United
States Naval Observatory, and the University of Washington.

\appendix

\section{Appendix}
In this appendix we provide a more detailed description of the
semisupervised mixture model we used in this study.  We describe what
needs to be included in the input data and how this differs from a
standard mixture model.  We explain the various parameters used in the
mixture model and how they make up the likelihood function that is
maximized to find the best model parameters.  We then describe how the
model is used to infer class membership for each of the objects in the
data set.  We explain how the Bayesian Information Criterion is used
to find the best overall model among available models. Finally, we
discuss how the model scales with the number of parameters, components
and objects in the data set.

\subsection{Data Scenario}
We first consider a data set that contains two types of samples: those
that contain a class label and those that are missing a class label,
called labeled and unlabeled respectively.  Each data point is
described by a feature vector $x_i \equiv
(x_{i1},x_{i2},\ldots,x_{id})$.  The class labels are drawn from a
finite set of known classes ${\cal P}_c$.  This description follows
that of \citet{Miller03a, Miller03b}.  In particular, we note that if
a sample is labeled then it must originate from one of the known
classes.  However, if the sample is unlabeled it may come from a new
class (one that is not in the set of known classes) or it may be an
unlabeled sample from a known class.  Moreover,
in the unlabeled case,
if it did come from a known class, it is uncertain which known class.

Mixture modeling is based on the premise that similar types of objects
tend to cluster together in feature space.  Thus, if a cluster
contains a mixture of labeled and unlabeled objects where the labels
are randomly missing, these unlabeled objects probably belong to the
class associated with the labeled data.  Furthermore, if a cluster
contains mainly unlabeled objects then it is possible that this cluster
describes a new class, one not contained in ${\cal P}_c$.  Clearly, the
presence or absence of a class label on a sample may be helpful in
distinguishing known classes from unknown.  Using this insight,
\citet{Miller03a, Miller03b} suggested using the presence or absence
of a class label as additional data which the mixture model must
explain.  Thus, the data set can be completely described as ${\cal
X}_m = \{{\cal X}_l,{\cal X}_u\}$, where now ${\cal X}_l = \{(x_1,{\rm
l},c_1), (x_2,{\rm l },c_2), \ldots, (x_{N_l},{\rm l},c_{N_l})\}$ is
the labeled data set and ${\cal X}_u = \{(x_{N_l+1}, {\rm m}), \ldots,
(x_{ N},{\rm m})\}$ is the unlabeled data set.  Here we use the new
random observation ${\cal L } \in \{{\rm l},{\rm m}\}$ that takes on
values indicating a sample is either labeled or missing the label.

\subsection{The Mixture Model}
Using the entire data set ${\cal X}_m$, \citet{Miller03a, Miller03b}
proposed a special mixture model to explain all the data, including
the presence or absence of a label for each sample.  This mixture
model included two types of mixture components which differed in the
way they generated the data values ${l}$ or ${m}$ indicating the
presence or absence of a label for each sample.  Predefined components
generate both labeled and unlabeled data from known classes, with the class
labels missing at random.  Nonpredefined components generate only
unlabeled data and may represent unknown or new classes: they capture
isolated clusters of unlabeled data.  We note that these two types of
components correspond directly to the data scenario described above.
The data from known classes have labels missing at random and are
described by predefined components.  The data from unknown classes are
purely unlabeled and are described by nonpredefined components.

Next we provide a more detailed description of the mixture model we
used, again following \citet{Miller03a, Miller03b}.  We begin with the
definitions of the relevant parameters in the model.  Our mixture
model consists of $k$ mixture components, denoted by ${\cal M}_k$,
$k=1,\ldots,M$. In this set of components there is a subset that are
predefined, ${\cal C}_{\rm pre}$ and a subset that are nonpredefined,
denoted ${\bar{\cal C}}_{\rm pre}$. We define $C \in {\cal P}_c \equiv
\{1,2,\ldots,N_c\}$ to be a random variable over the $N_c$ known
classes, with $c(x) \in {\cal P}_c$ the class label for sample $x$.
The prior probability for component $k$ is denoted $\alpha_k$.  We
denote by $\theta_k$ the parameter set specifying component $k$'s
component-conditional joint feature density, and let $f(\underline{x}
| \theta_k )$ denote this density.  It is useful to introduce a new
class set $\tilde{\cal P}_c = \{1,2,\ldots,N_c,u\}$. $\tilde{\cal
P}_c$ augments the original class set ${\cal P}_c$ by adding the value
$u$ which is used to indicate that a sample is unlabeled. With respect
to the augmented set $\tilde{\cal P}_c$ every sample is now
``labeled'', with the unlabeled samples taking on the class label
``u''.  We assume each model component has a different random
label generator which we write as ${\rm Prob}[c|{\cal M}_k] \equiv
\beta_{c|k}$. Here the class $c$ is selected from the augmented class
set: $c \in \tilde{\cal P}_c$.  Note that $\sum\limits_{c \in
\tilde{\cal P}_c} \beta_{c|k} = 1$.  The function $\beta_{c|k}$
measures the fraction of samples from component $k$ that belong to
class $c$.  In particular, $\beta_{u|k}$ is the fraction of unlabeled
samples from component $k$.  Strictly speaking, for a nonpredefined
component $\beta_{u|k} = 1$, i.e., all samples from the nonpredefined
component, $k$, are unlabeled.  In practice, we take $\beta_{u|k}
\simeq 1$; see discussion below.  In summary, the mixture model is
based on the parameter set $\Lambda = \{\{\alpha_k\},\{\theta_k\},
\{\beta_{c|k}\}\}$.

{\em Hypothesis for Random Generation of the Data}---The model of
\citet{Miller03b} hypothesizes that each sample from ${\cal X}_m$ is
generated independently, based on the parameter set $\Lambda$,
according to the following stochastic generation process:

\begin{enumerate}
\item Randomly select a component ${\cal M}_j$ according to prior
probability $\{\alpha_k\}$.

\item Randomly select a sample $x$ according to $P(x | \theta_j)$ and a
label $c$ according to $\{\beta_{c|k}\}$.

\end{enumerate}

{\em Joint Data Likelihood}---The log of the joint data likelihood
associated with this model is
\begin{eqnarray}
\label{newlik}
{\cal L} = \sum\limits_{x \in {\cal X}_m}
\log\sum\limits_{k =1}^M \alpha_k f(x | \theta_k)
\beta_{c(x) | k},
\end{eqnarray}
where $c(x) \in \tilde{\cal P}_c$.  The model parameters $\Lambda$ can
be chosen to maximize the log-likelihood, Equation \ref{newlik}, via
the Expectation-Maximization (EM) algorithm (e.g. \citet{Duda}).

This model does not explicitly discover new class components, i.e.,
mixture components that are purely unlabeled.  However, suppose that
for a given component ${\cal M}_j$ two situations pertain. First
suppose that $\beta_{u | j} \simeq 1$.  This means that almost all of
the samples owned by this component are unlabeled.  Furthermore,
suppose that $\beta_{u|j}$ is also significantly greater than the
average value $\frac{1}{M} \sum\limits_{j'=1}^M \beta_{u|j'}$, meaning
that the number of unlabeled samples owned by this component is larger
than for most other components.  Then, for component ${\cal M}_j$ the
fraction of unlabeled data that it owns is unusually high and we
categorize this component as ``nonpredefined'', i.e. ${\cal M}_j \in
{\bar{\cal C}}_{\rm pre}$.  Such components describe a class of
objects that are putatively unknown, novel, or new.  All other
components are categorized as ``predefined'', and represent known
class data.  In other words we use the following strategy for new
class discovery in a mixed labeled and unlabeled data scenario: (1)
learn a mixture model to maximize the log likelihood, Equation
\ref{newlik}; (2) for each component, declare it ``nonpredefined'' if
$\beta_{u | j} - (\frac{1}{M} \sum\limits_{j'=1}^M \beta_{u|j'}) >
\delta$; otherwise, declare it ``predefined''.  Here, $\delta$ is a
suitably chosen threshold.  In practice, we declare a component
``nonpredefined'' when its value $\beta_{u|j}$ is closer to $1.0$ than
to the average value, i.e., we choose $\delta = \frac{1}{2}(1 -
\frac{1}{M}\sum\limits_{j'=1}^M \beta_{u|j'})$.  We have found this
choice for $\delta$ to give reasonable results for a variety of
experimental conditions (for different data sets and for different
fractions of labeled data).

\subsection{Posterior Probabilities for Statistical Inference}
Using the procedure just described we produce the maximum likelihood
model based on the optimal parameter set defined by $\Lambda$.  This
model can be used for the two inference tasks that are of interest to
us:  1) Classification of a given sample to one of the known classes
and 2) discrimination between known and unknown classes.  For a given
sample $x$, we can perform classification using the {\it a posteriori}
probabilities 

\begin{eqnarray}
\label{inf1}
P[C=c | \underline{x}; \Lambda] = \frac{\sum\limits_{k \in {\cal
C}_{\rm pre}} \alpha_k f(\underline{x} | \theta_k)
(\frac{\beta_{c|k}}{1 - \beta_{u|k}})}{\sum\limits_{k \in {\cal C
}_{\rm pre}} \alpha_k f(\underline{x} | \theta_k)
}, c \in {\cal P}_c.
\end{eqnarray}

These can be used in a maximum {\it a posteriori} (MAP) class decision rule,
i.e. $c^{\ast} = \arg\max\limits_c P[C=c | \underline{x}; \Lambda]$. 

To handle the second inference task, discrimination between the
hypotheses that an unlabeled sample originates from a known versus an
unknown class, we need the {\it a posteriori} probability that the
given feature vector is generated by a nonpredefined component.  This
is given by
\begin{eqnarray}
\label{inf2}
P[{\cal M}_{\rm np} | \underline{x} \in {\cal X}_u] =
 \frac{\sum\limits_{k \in {\bar{\cal C}}_{\rm pre}} \alpha_k
 f(\underline {x} | \theta_k) \beta_{u|k}}{\sum\limits_k \alpha_k
 f(\underline{x} | \theta_k) \beta_{u|k}}.
\end{eqnarray}

If $P[{\cal M}_{\rm np} | \underline{x} \in {\cal X}_u] >
\frac{1}{2}$, then the sample is declared to belong to an unknown
class; otherwise it is declared as a known class sample.

\subsection{Model Order Selection}
The expectation-maximization learning approach we used to find the
maximum likelihood model parameters assumes that the number of mixture
components $M$ (the model order) is fixed and known.  However, in
practice this size must be estimated.  Model order selection is a
difficult and pervasive problem, with several criteria proposed
\citep{Schwarz, Wallace, Mclachlan} and no consensus on the right one.
When attempting class discovery, as we are doing here, accurate model
order selection is critical.  Specifically, the nonpredefined
components in the validated solution will be taken as candidates for
new classes.  These new classes will be examined by a domain expert to
determine their validity as new classes, prompting additional study as
needed.  Accurate model order selection is thus paramount for
successful new class discovery.  As discussed in the main text, and in
\citet{Miller03a}, and \citet{Miller03b} we use the Bayesian Information
Criterion (BIC) \citep{Schwarz} to decide between models with
different numbers of components.  The BIC model selection criterion is
written in the form
\begin{eqnarray}
\label{BIC}
BIC(M) = \frac{N_p(M)}{2} \log N - {\cal L}, 
\end{eqnarray}
with $N_p(M)$ the number of free parameters in the $M$-component
mixture model and $N$ the data length.  The first term is the penalty
on model complexity, with the second term the negative log-likelihood.
We applied BIC in a ``wrapper-based'' model selection approach; i.e.,
we built models for increasing $M$, evaluated BIC for each model, and
then selected the model with minimum BIC cost (see Figure 1 in the
main text).

\subsection{Computational Complexity}
The computational complexity of our (EM-based) learning
is $O(CDNI)$, with $C$ the number of components, $D$ the
data dimensionality, $N$ the number of data points, and $I$
the number of learning iterations before the EM algorithm satisfies
the convergence criterion (based on diminishing relative gain in log-likelihood
from one iteration to the next).
While the number of learning iterations required to converge may in general
depend upon the number of components and data dimensions, we have found experimentally that the total
learning time does grow approximately linearly in these variables.


\begin{thebibliography}{}

\bibitem[Akaike (1973)]{Akaike} Akaike, H. 1973, in B. N. Petrov
  \& F. Csaki, eds. Second Int'l Symp. on Info. Theory, Budapest,
  p. 267

\bibitem[Bailer-Jones et al.(1997)]{Bailer-Jones97} Bailer-Jones,
C.~A.~L., Irwin, M., Gilmore, G., \& von Hippel, T.\ 1997, \mnras,
292, 157

\bibitem[Bailer-Jones et al.(1998)]{Bailer-Jones98} Bailer-Jones, 
C.~A.~L., Irwin, M., \& von Hippel, T.\ 1998, \mnras, 298, 361 
 
\bibitem[Ball et al.(2004)]{Ball04} Ball, N.~M., Loveday, J.,
Fukugita, M., Nakamura, O., Okamura, S., Brinkmann, J., \& Brunner,
R.~J.\ 2004, \mnras, 348, 1038

\bibitem[Banfield \& Raftery (1993)]{Raftery} Banfield J. D. \&
Raftery, A. E. 1993, Biometrics, 39, 803

\bibitem[Basu et al. (2002)]{Basu1} Basu, S., Banerjee, A., \& Mooney,
 R. 2002, Intl. Conf. on Machine Learning, 1, 19

\bibitem[Bazell \& Aha(2001)]{Bazell01} Bazell, D., \& Aha, D.~W.\
2001, \apj, 548, 219

\bibitem[Bazell \& Miller (2005)]{Bazell05} Bazell, D. \& Miller,
  D.J. 2005, \apj, 618, 723

\bibitem[Connolly et al.(1995)]{Connolly95} Connolly, A.~J., Szalay,
A.~S., Bershady, M.~A., Kinney, A.~L., \& Calzetti, D.\ 1995, \aj,
110, 1071

\bibitem[Connolly \& Szalay(1999)]{Connolly99} Connolly, A.~J., \&
Szalay, A.~S.\ 1999, \aj, 117, 2052

\bibitem[Duda, Hart, \& Stork (2001)] {Duda} Duda, R. O., Hart, P. E.,
\& Stork, D. G. 2000, Pattern Classification (2nd ed.; New York: Wiley)

\bibitem[Graham \& Miller (2006)]{Graham06} Graham, M. W. \& Miller,
  D. J. 2006, IEEE Trans. Sig. Proc., 54, 1289

\bibitem[Hall, et al. (2002)]{Hall02} Hall, P. B., et al., 2002,
  \apjs, 141, 267

\bibitem[Jarrett et al.(2000)]{Jarrett00} Jarrett, T.~H., 
Chester, T., Cutri, R., Schneider, S., Skrutskie, M., \& Huchra, J.~P.\ 
2000, \aj, 119, 2498 

\bibitem[Kelly \& McKay (2004)]{Kelly04} Kelly, B. C. \& McKay,
  T. A. 2004, \aj, 127, 625

\bibitem[Kelly \& McKay (2005)]{Kelly05} Kelly, B. C. \& McKay,
  T. A. 2005, \aj, 129, 1287

\bibitem[Kleinman et al.(2004)]{Kleinman04} Kleinman, S.~J., et al.\
2004, \apj, 607, 426

\bibitem[Kohonen(2001)]{Kohonen01} Kohonen, T.\ 2001, 
Self-organizing maps.~3rd ed.~Berlin: Springer, 2001, 501 p.~Springer 
series in information sciences  

\bibitem[M\"{a}h\"{o}nen \& Hakala(1995)]{Mahonen95} M\"{a}h\"{o}nen,
P.~H., \& Hakala, P.~J.\ 1995, \apjl, 452, L77

\bibitem[Mclachlan \& Peel (2000)]{Mclachlan} McLachlan, G., \& Peel,
D. 2000, Finite Mixture Models, (New York: Wiley)

\bibitem[Miller \& Coe(1996)]{Miller96} Miller, A.~S., \& Coe, M.~J.\
1996, \mnras, 279, 293

\bibitem[Miller \& Browning (2003a)]{Miller03a} Miller D. J. \& Browning,
J. 2003, IEEE Trans. on Pattern Anal. and Machine Intell., 25,
1468

\bibitem[Miller \& Browning (2003b)]{Miller03b} Miller D. J. \&
Browning, J. 2003, IEEE Intl. Workshop on Neural Networks for
Signal Processing, Toulouse, France, 1, 489

\bibitem[Miller \& Uyar (1997)]{Miller97} Miller D. J. \& Uyar,
H. 1997, Neural Information Processing Systems, 9, 571

\bibitem[Naim et al.(1995a)]{Naim95a} Naim, A., et al.\ 1995, 
\mnras, 274, 1107 

\bibitem[Naim et al.(1995b)]{Naim95b} Naim, A., Lahav, O., Sodr\'{e}
Jr., L., \& Storrie-Lombardi, M. C. 1995, \mnras, 275, 567

\bibitem[Naim et al.(1997)]{Naim97a} Naim, A., Ratnatunga, 
K.~U., \& Griffiths, R.~E.\ 1997, \apjs, 111, 357 

\bibitem[Nichol et al.(2001)]{Nichol01} Nichol, R.~C., et al.\ 2001,
Mining the Sky, 613
 
\bibitem[Nigam et al. (2000)]{Nigam} Nigam K., McCallum, A., Thrun,
  S., \& Mitchell, T., 2000, Machine Learning, 39, 1

\bibitem[Odewahn et al.(1992)]{Odewahn92} Odewahn, S. C., Stockwell,
E. B., Pennington, R. L., Humphreys, R. M., \& Zumach, W. A.  1992,
\aj, 103, 318

\bibitem[Owens et al. (1996)]{Owens96} Owens E. A., Griffiths, R. E., \&
  Ratnatunga, K. U. 1996, \mnras, 281, 153.

\bibitem[Rajaniemi \& M\"{a}h\"{o}nen(2002)]{Rajaniemi02} Rajaniemi,
H.~J. \& M\"{a}h\"{o}nen, P. 2002, \apj, 566, 202

\bibitem[Richards, et al. (2002)]{Richards02} Richards, G. T., et
  al. 2002, \aj, 123, 2945

\bibitem[Richards, et al. (2004)]{Richards04} Richards, G. T., et al.,
  2004, \apjs, 155, 257

\bibitem[Seibert, et al. (2005)]{Seibert05} Seibert, M., et al., 2005,
  \apjl, 619, L23

\bibitem[Schwarz (1978)]{Schwarz} Schwarz G. 1978, Annals of
Statistics, 6, 461

\bibitem[Shashahani \& Landgrebe (1994)]{Shashahani} Shashahani B. \&
Landgrebe D. 1994, IEEE Trans. on Geo. and Remote Sens., 32, 1087

\bibitem[Storrie-Lombardi, et al.(1992)]{Storrie92} Storrie-Lombardi
M. C., Lahav O., Sodr\'{e} Jr. L., \& Storrie-Lombardi L. J. 1992,
\mnras, 259, 8

\bibitem[Stoughton, et al. (2002)]{Stoughton} Stoughton, C. et
  al. 2002, \aj, 123, 485

\bibitem[Strateva, et al. (2001)]{Strateva01} Strateva, I., et
  al. 2001, \aj, 122, 1861

\bibitem[Suchkov, Hanisch, \& Margon (2005)]{Suchkov05} Suchkov,
  A. A., Hanisch, R. J., \& Margon B. 2005 astro-ph/0508501

\bibitem[von Hipple et al.(1994)]{Vonhipple94} von Hipple,
T., Storrie-Lombardi, L., Storrie-Lombardi, M. C., \& Irwin, M. 1994,
\mnras, 269, 97

\bibitem[Wallace \& Freeman (1987)]{Wallace} Wallace C. S.,\& Freeman
P. R. 1987, Journal of the Royal Statistical Society, B, 49, 223

\bibitem[Weir et al.(1995)]{Weir95a} Weir, N., Fayyad, U.~M., 
\& Djorgovski, S.\ 1995, \aj, 109, 2401 

\bibitem[Willemsen et al.(2005)]{Willemsen05} Willemsen, P.~G.,
Hilker, M., Kayser, A., \& Bailer-Jones, C.~A.~L.\ 2005, \aap, 436,
379

\bibitem[Yip et al. (2004)]{Yip04} Yip, C.W., et al. 2004, \aj, 128,
  585

\end{thebibliography}
\end{document}